\documentclass[a4paper]{panl}
\usepackage{graphicx}
\usepackage{amssymb}
\usepackage{amsfonts}
\usepackage{amsmath}
\usepackage{longtable}
\usepackage{lscape}
\usepackage{epsfig}
\usepackage{multirow}

\originalTeX
\begin{document}

\title{Towards the structure of a cubic interaction vertex for \\ massless integer  higher spin fields
}
\maketitle
\setcounter{footnote}{0} \authors{A.A.\,Reshetnyak$^{a,b,c,}$\footnote{E-mail: reshet@tspu.edu.ru}}
\from{$^{a}$\, Center for Theoretical Physics, Tomsk State Pedagogical University, Tomsk, Russia}
\from{$^{b}$\, National Research Tomsk Polytechnic University, Tomsk, Russia}
\from{$^{c}$\, National Research Tomsk State University, Tomsk, Russia}
\begin{abstract}

\vspace{0.2cm}
The structure of  a cubic Lagrangian vertex is clarified for irreducible fields of  helicities $s_1, s_2, s_3$ in a $d$-dimensional Minkowski space.  An explicit form of the operator $\mathcal{Z}_j$ entering the vertex in a non-multiplicative way (examined in \cite{BRcub} for $j=1$) is obtained. The solution is found within the  BRST approach with complete  BRST operators, which contain all constraints corresponding to the conditions that extract the irreducible fields, including trace operators.
\end{abstract}
\vspace*{6pt}


\label{sec:intro}
\section*{Introduction}

The theory of interacting higher-spin fields   has become one of the topical areas of theoretical
and mathematical high-energy physics (for a review, we can recommend, e.g., \cite{reviews},
\cite{revBCIV}, \cite{reviews3}, \cite{rev_Bekaert}, \cite{reviewsV}). It is anticipated that
interacting higher-spin fields will open up new opportunities in the search for elementary particles beyond
the Standard Model, and will also contribute to the emergence of pioneering approaches
to the unification of fundamental interactions.

In our recent paper \cite{BRcub}, a general Lagrangian cubic vertex has been obtained for unconstrained interacting
fields with integer-helicities  in Minkowski spaces  (see \cite{Manvelyan}, \cite{Manvelyan1}, \cite{Joung}, \cite{frame-like1},
\cite{Metsaev0512}, \cite{Metsaev0712}, \cite{BRST-BV3}, \cite{frame-like2}, \cite{BKTW}
for the study of cubic vertices in different approaches). In contrast to the previously known results on cubic vertices,
the study of \cite{BRcub} does not impose on interacting fields any algebraic relations that do not follow from
the least action principle.
The vertex is derived based on a BRST-closed solution of an operator equation arising from the condition that
demands the preservation of gauge invariance for a deformed free action with respect to deformed gauge transformations,
which, in their turn, follow from the application of an unconstrained BRST approach (developed, for example, in \cite{PT}, \cite{BPT}, \cite{BR}, \cite{BFPT}, \cite{reviews3}; for the equivalence of the constrained \cite{BRST-BV3} and unconstrained BRST approaches, see \cite{Reshetnyak_con})  to the Lagrangian description of  higher-spin free field models in Minkowski and anti-de Sitter spaces. The found vertex corresponds to the cubic vertex \cite{Metsaev0512} deduced using the light-cone formalism in terms of physical degrees of freedom, and preserves the irreducibility of a representation for interacting fields, in particular, the number of physical degrees of freedom under a deformation of a free Lagrangian formulation.

The vertex $|{V}{}^{(3)}\rangle_{(s)_3}$ found in \cite{BRcub} (see (\ref{genvertex}),
(\ref{Vmets})  for the definition) contains operator quantities including trace $U^{(s_i)}_{j_i}$, differential $\mathcal{L}^{(i)}_{k_i} $ operators entering multiplicatively and corresponding to the spin values $s_i$, $i=1,2,3$, as well as some operators $\mathcal{Z}_j$ characterized simultaneously by three sets of spins, $s_1, s_2, s_3$. An expression for the operator $\mathcal{Z}_1$ has been found in \cite{BRcub}. The present article is aimed to finding an explicit representation for the operator $\mathcal{Z}_j$ entering the vertex non-multiplicatively for $j=2,3,...$.

The paper has the following organization. In Section~\ref{sec:preparation}, the results of the BRST construction involving a complete BRST operator are presented as applied to deriving a cubic vertex for unconstrained fields of integer helicities, $s_1$, $s_2$, $s_3$. In Section~\ref{sec:BRSTinterZ}, we obtain
the operators $\mathcal{Z}_j$ for $j>1$. Conclusion summarizes the results.


We use the conventions of \cite{BRcub}: $\eta_{\mu\nu} = diag (+, -,...,-)$ for a metric tensor with Lorentz indices $\mu, \nu = 0 ,1,...,d-1$,
the notation $\epsilon(F)$,$gh(F)$,$[F,\,G\}$, $[x]$ for the respective Grassmann parity and ghost number of a homogeneous quantity $F$,
as well as the supercommutator of quantities $F,G$ and the integer part of a number $x$.

\section{BRST approach to a cubic interaction vertex}
\label{sec:preparation}

A Lagrangian formulation for a cubic vertex within the BRST approach to interacting real-valued totally symmetric massless fields $\phi^{(i)}_{\mu(s_i)}$ $\equiv$ $\phi^{(i)}_{\mu_1...\mu_{s_i}}(x)$, $i=1,2,3$ with integer higher helicities $s_1$, $s_2$, $s_3$
in a $d$-dimensional Minkowski space determines a gauge theory of first-stage reducibility in a configuration space
$\mathcal{M}^{(s)_3}_{cl}$ \cite{BRcub} with the action functional
\begin{eqnarray}\label{S[n]}
  && \hspace{-1.0em}S_{[1]|(s)_3}[\chi]  =  \sum_{i=1}^{3} \mathcal{S}_{0|s_i}[\chi^{(i)}_{s_i}]   +
  g  \int \prod_{e=1}^{3} d\eta^{(e)}_0  \Big( {}_{s_{e}}\langle \chi^{(e)} K^{(e)}
  \big|  V^{(3)}\rangle_{(s)_{3}}+h.c. \Big)\hspace{-0.25ex}, \\
\label{PhysStatetot} &&  \hspace{-0.5em} \mathcal{S}_{0|s_i}[\chi^{(i)}_{s_i}] =  \mathcal{S}_{0|s_i}[\phi^{(i)},\phi^{(i)}_1,...]=
  \int d\eta^{(i)}_0 {}_{s_i}\langle\chi^{(i)}|
K^{(i)}Q^{(i)}|\chi^{(i)}\rangle_{s_i},
\end{eqnarray}
being invariant up to the first order in the interaction constant $g$ with respect to non-Abelian gauge transformations with zero-level parameters
$\big| \Lambda^{(i)} \rangle_{s_i}$
\begin{eqnarray}
  && \delta_{[1]} \big| \chi^{(i)} \rangle_{s_i}  =  Q^{(i)} \big| \Lambda^{(i)} \rangle_{s_i} -
g \int \prod_{e=1}^{2} d\eta^{(i+e)}_0  \Big( {}_{s_{i+1}}\langle
\Lambda^{({i+1})}K^{(i+1)}\big|\otimes   \label{cubgtr}\\
   &&
\ \   \phantom{\delta_{[1]} \big| \chi^{(i)} \rangle_{s_i}} \otimes{}_{s_{i+2}}
   \langle \chi^{({i+2})}K^{(i+2)}\big|  +(i+1 \leftrightarrow i+2)\Big)
\big|{V}{}^{(3)}\rangle_{(s)_{3}}, \nonumber
\end{eqnarray}
which are invariant, with the same accuracy, under gauge transformations with independent parameters,
$\big| \Lambda^{(i)1} \rangle_{s_i}$
\begin{eqnarray}
&& \delta_{[1]} \big| \Lambda^{(i)} \rangle_{s_i}  =  Q^{(i)} \big| \Lambda^{(i)1} \rangle_{s_i}
 -g  \int \prod_{e=1}^{2} d\eta^{(i+e)}_0  \Big( {}_{s_{i+1}}\langle \Lambda^{(i+1)1}K^{(i+1)}\big|\otimes    \label{cubggtr}\\
   &&
\ \   \phantom{\delta_{[1]} \big| \chi^{(i)} \rangle_{s_i}} \otimes {}_{s_{i+2}}
\langle \chi^{({i+2})}K^{(i+2)}\big| +(i+1
\leftrightarrow i+2)\Big) \big|{V}{}^{(3)}\rangle_{(s)_{3}}.
\nonumber
\end{eqnarray}
In (\ref{S[n]}), (\ref{cubgtr}), (\ref{cubggtr}), the conventions $(s)_{3}
\equiv (s_1,s_2,s_3)$, $[i+3 \simeq i]$, as well as  $S_{[1]}= \sum_{i=1}^{3} \mathcal{S}_{0| s_i}+S_{1}$, and
 $\delta_{[1]}=\delta_{0}+\delta_{1}$  for the deformed action and gauge transformations are used.
 The functional $\mathcal{S}_{0|s_i}$ (\ref{PhysStatetot}) is an action being quadratic in the fields for the $i$-th copy of
 a set of fields $|\chi^{(i)}\rangle_{s_i}$. The space $\mathcal{M}^{(s)_3}_{cl}$ is parametrized by basic fields $\phi^{(i)}_{\mu(s)_i}$ and sets of auxiliary fields $\phi^{ (i)}_{1\mu({s_i}-1)},...$ of smaller rank, embedded in the vectors $|\chi^{(i)}\rangle_{s_i}$ of a Hilbert space, $\mathcal{H }^{(i)}_{tot}$ $=$ $\mathcal{H}^{(i)}\otimes \mathcal{H}^{(i)}{}'\otimes \mathcal{H}^{ (i)}_{gh}$, $i=1,2,3$
\begin{eqnarray}
\hspace{-1em}&& |\chi^{(i)}\rangle_{s_i}  =
|\Phi^{(i)}\rangle_{s_i}+\eta^{(i)+}_1\Big(\mathcal{P}_1^{(i)+}|\phi^{(i)}_2\rangle_{s_i-2} +\eta_{11}^{(i)+}\mathcal{P}_1^{(i)+}\mathcal{P}_{11}^{(i)+}|\phi^{(i)}_{22}\rangle_{s_i-6} \label{spinctotsym} \\
\hspace{-1em}&\hspace{-1em}&\hspace{-1em} \phantom{ |\chi^0_c} +\mathcal{P}_{11}^{(i)+}|\phi^{(i)}_{21}\rangle_{{s_i}-3}\Big) +\eta_{11}^{(i)+}\Big(\mathcal{P}_1^{(i)+}|\phi^{(i)}_{31}\rangle_{s_i-3}+\mathcal{P}_{11}^{(i)+}|\phi^{(i)}_{32}\rangle_{s_i-4}\Big)+  \eta_0^{(i)}\Big(\mathcal{P}_1^{(i)+} \nonumber \\ \hspace{-1em}&\hspace{-1em}&\hspace{-1em} \phantom{ |\chi^0_c}  \times |\phi^{(i)}_1\rangle_{s-1} +\mathcal{P}_{11}^{(i)+}|\phi^{(i)}_{11}\rangle_{s-2}   +  \mathcal{P}_1^{(i)+}\mathcal{P}_{11}^{(i)+}\Big[ \eta^{(i)+}_{1} |\phi^{(i)}_{12}\rangle_{s_i-4}+\eta^{(i)+}_{11} |\phi^{(i)}_{13}\rangle_{s_i-5}\Big]\Big).\nonumber
\end{eqnarray}
The quantities $\eta^{(i)}_0$, $\eta_1^{(i)+}$, $\eta_{11}^{(i)+}$, $\mathcal{P}_{1} ^{(i)+}$, $\mathcal{P}_{11}^{(i)+}$ are ghost operators generating Hilbert spaces $\mathcal{H}^{(i)}_{gh} $ with ghost-independent vectors $|\phi^{(i)}_{..}\rangle_{s-...}$. $Q^{(i)}$ and $K^{(i)}$ in (\ref{S[n]})-(\ref{cubggtr}) stand for the BRST operator and the operator defining an inner product in the space $\mathcal{H}^{(i)}_{tot}$; the index $s_i$ determines the spin value of the corresponding vector. Vector gauge parameters of zero $|\Lambda^{(i)}\rangle_{s_i}$ and first  $|\Lambda^{1(i)}\rangle_{s_i}$ levels,
\begin{eqnarray}
\hspace{-1em}&\hspace{-1em}&\hspace{-1em} |\Lambda^{(i)}\rangle_{s_i}  =  \mathcal{P}_1^{(i)+}  |\xi^{(i)}\rangle_{s_i-1}+\mathcal{P}_{11}^{(i)+}|\xi_{1}^{(i)}\rangle_{s_i-2} +\mathcal{P}_1^{(i)+}\mathcal{P}_{11}^{(i)+}\Big(\eta_1^{(i)+}|\xi_{11}^{(i)}\rangle_{s_i-4} \label{parctotsym}
\\
\hspace{-1em}&\hspace{-1em}&\hspace{-1em} \phantom{|\chi^1\rangle_s}
 + \eta_{11}^{(i)+}|\xi_{12}^{(i)}\rangle_{s_i-5}+  \eta^{(i)}_0|\xi_{01}^{(i)}\rangle_{s_i-3}\Big)  ,
  \nonumber\\
\hspace{-1em}&\hspace{-1em}&\hspace{-1em} {|\Lambda^{1(i)}\rangle_{s_i}}  =
\mathcal{P}_1^{(i)+}\mathcal{P}_{11}^{(i)+}|\xi^{1(i)}\rangle_{s_i-3},
\end{eqnarray}
as elements of respective $Q^{(i)}$-complexes (see, for example, \cite{BPT},\cite{BR}) determine a distribution of Grassmann parity and ghost number for $\hspace{-0.1ex}|\chi^{(i)}\rangle_{s_i}$, $\hspace{-0.1ex}|\Lambda^{(i)}\rangle_{s_i}$, $\hspace{-0.1ex}|\Lambda^{1(i)}\rangle_{s_i}$, respectively, $(0,0)$, $(1,-1),$ $(0,-2)$.

Unitary massless irreducible representations of the Poincaré  $ISO(1, d-1)$ group with integer helicities $(s)_{3}$ are realized on basic fields in the free approximation ($g=0$) described by d'Alembert equations, as well as by conditions of no divergence and tracelessness \cite{Wigner}, equivalently represented by non-Lagrangian operator conditions for the vector $|\phi^{(i)}\rangle \in$ $\mathcal{H}^{(i)}$, $i=1,2 .3$
\begin{eqnarray}\label{irrepint}
     &&       \big(l^{(i)}_0,\, l^{(i)}_1,\, l^{(i)}_{11}, g^{(i)}_0 -d/2\big)|\phi^{(i)}\rangle  = (0,0,0,s_i)|\phi^{(i)}\rangle. \\
     \label{FVoper}
&&   |\phi^{(i)}\rangle  =  \sum_{s_i\geq 0}\frac{\imath^{s_i}}{s_i!}\phi^{{(i)}\mu(s_i)}\prod_{j=1}^{s_i} a^{(i)+}_{\mu_j}|0\rangle, \\
&&   \big(l^{(i)}_0,\, l^{(i)}_1,\, l^{(i)}_{11}, g^{(i)}_0\big) = \big(\partial^{(i)\nu}\partial^{(i)}_\nu ,\, - \imath a^{(i)\nu}  \partial^{(i)}_\nu ,\, \frac{1}{2}a^{(i)\mu} a^{(i)}_\mu ,  -\frac{1}{2}\big\{a^{(i)+}_{\mu},\, a^{(i)\mu}\big\}\big).\nonumber
\end{eqnarray}
The operators $l^{(i)}_0,\, l^{(i)}_1,\, l^{(i)}_{11}, g^{(i)}_0$ and the basic vector $|\phi^{(i)}\rangle$ are defined in a Fock space $\mathcal{H}^{(i)}$
generated by bosonic oscillators $a^{(i)}_\mu, a^{(i)+}_\nu$ \- ($[a^{(i)}_\mu,\hspace{-0.1ex} a^{(i)+}_\nu]$ $=$ $- \eta_{\mu\nu}$). The basic vector $|\phi^{(i)}\rangle_{s_i}$ is embedded in the vector $|\Phi^{(i)}\rangle_{s_i}$, depending, along with the remaining ones, $|\phi^{(i)}_{..}\rangle_{s-...}$,
also on auxiliary bosonic oscillators $ b^{(i) +}$ ($[b^{(i)},\, b^{(j)+}]=\delta^{ij}$ which form a basis in the Fock space $\mathcal{H}^{(i)}{}'$.

Each of the BRST operators  $Q^{(i)}$ ($(\epsilon, gh) Q^{(i)} = (1, 1)$) is constructed using
a corresponding system of constraints: $l^{(i)}_0,\, l^{(i)}_1,\, l^{(i)+}_1,\,
l^{(i)}_{11},\, l^{(i)+}_{11}=\frac{1}{2}a^{(i)+\nu}a^{(i)+}_{\nu}$ and contains anticommuting ghost operators, $\eta^{(i)}_0$, $\eta_1^{(i)+}$, $\eta^{(i)}_1$,
$\eta_{11}^{(i)+}$, $\eta^{(i)}_{11}$, ${\cal{}P}^{(i)}_0$, $\mathcal{P}^{(i)}_{1}$,
$\mathcal{P}^{(i)+}_{1},$ $\mathcal{P}^{(i)}_{11}$, $\mathcal{P}^{(i)+}_{11},$
\begin{eqnarray}
\hspace{-0.9em}&\hspace{-0.9em}&\hspace{-0.9em} {Q}^{(i)} =
\eta^{(i)}_0l^{(i)}_0+\eta_1^{(i)+}l_1^{(i)}+l_1^{(i)+}\eta_1^{(i)}+
\eta_{11}^{(i)+}\widehat{L}{}^{(i)}_{11}+\widehat{L}{}_{11}^{(i)+}\eta_{11}^{(i)} +
{\imath}\eta_1^{(i)+}\eta_1^{(i)}{\cal{}P}{}^{(i)}_0
\label{Qctotsym},
\end{eqnarray}
where BRST-extended traceless constraints have the form
\begin{eqnarray}
\hspace{-0.5ex}&\hspace{-0.5ex}&\hspace{-0.5ex}
\big(\widehat{L}{}^{(i)}_{11} ,\,\widehat{L}{}^{(i)+}_{11}\big) =  \big(
L^{(i)}_{11}+\eta^{(i)}_{1} \mathcal{P}^{(i)}_{1} , \,
L^{(i)+}_{11}+\mathcal{P}^{(i)+}_{1}\eta^{(i)+}_{1} \big).
\label{extconstsp2}
\end{eqnarray}
Here, the operators
\begin{eqnarray}\label{L11L11+}
L^{(i)}_{11}=l^{(i)}_{11}+(b^{(i)+}b^{(i)}+h^{(i)})b^{(i)},\,\, \ \  L^{(i)+}_{11}=l^{(i)+}_{11}+b^{(i)+}
\end{eqnarray}
depend on parameters $h^{(i)} = h^{(i)}(s_i)=-s_i - \frac{d-6}{2}$. Three sets of operators, $l^{(i)}_0$ ,$l^{(i)}_1$, $l^{(i)+}_1; L^{(i)}_{11}, L_{11}^{(i)+},
G^{(i)}_0$, commute with one another at $i\ne j$ and form 3 isometry subalgebras in Minkowski space and 3 subalgebras $so(1,2)$
\begin{equation}\label{subalgebr}
[l^{(i)}_0, l^{(i)(+)}_1] \hspace{-0.1em}=\hspace{-0.1em} 0, \, [l^{(i)}_1,l_1^{(i)+}]\hspace{-0.1em}= \hspace{-0.1em}l^{(i)}_0;  \ \    [L^{(i)}_{11},  L_{11}^{(i)+}] \hspace{-0.1em}=\hspace{-0.1em} G^{(i)}_0,\,
[G^{(i)}_0, L_{11}^{(i)+}]\hspace{-0.1em} =\hspace{-0.1em} 2 L_{11}^{(i)}
\end{equation}
with independent cross-commutators: $\hspace{-0.55em}$ $[l^{(i)}_1,\hspace{-0.1em}G^{(i)}_{0}]\hspace{-0.1em}=l^{(i)}_1\hspace{-0.1em}$, $[l^{(i)}_1,\hspace{-0.1em}L^{(i)+}_{11}]$ $=-l_1^{(i)+}$.

The ghost operators satisfy the non-vanishing anticommutation relations
\begin{eqnarray}\label{ghanticomm}
  && -\imath\{\eta^{(i)}_0, \mathcal{P}^{(j)}_0\}= \{\eta^{(i)}_1, \mathcal{P}_1^{(j)+}\}=\{\eta^{(i)+}_1, \mathcal{P}^{(j)}_1\}=\\
   && \{\eta^{(i)}_{11},   \mathcal{P}_{11}^{(j)+}\}=\{\eta_{11}^{(i)+},   \mathcal{P}^{(j)}_{11}\}=\delta^{ij}. \nonumber
\end{eqnarray}
The given theory is characterized by the spin operators
$\sigma^{(i)}$,
\begin{eqnarray}
&& \sigma^{(i)}  =   G^{(i)}_0+ \eta_1^{(i)+}\mathcal{P}^{(i)}_{1}
-\eta^{(i)}_1\mathcal{P}_{1}^{(i)+}  + 2(\eta_{11}^{(i)+}\mathcal{P}^{(i)+}_{11} -\eta^{(i)+}_{11}\mathcal{P}_{11}^{(i)+}) .
\label{extconstsp3}
\end{eqnarray}
Here, $G^{(i)}_0=g^{(i)}_0 +2b^{(i)+}b^{(i)}+h^{(i)}$ is a converted particle number operator in the Fock space $\mathcal{H}^{(i)}\otimes \mathcal{H}^{(i)}{}'$.
The operator $\sigma^{(i)}$ selects eigenvectors with a definite spin value $s_i$ in the space $\mathcal{H}^{(i)}_{tot}$
\begin{eqnarray}
 \hspace{-0.5ex}&\hspace{-0.5ex}&\hspace{-0.5ex} \sigma^{(i)} \big(|\chi^{(i)}\rangle_{s_i},\,
 |\Lambda^{(i)}\rangle_{s_i},\, |\Lambda^{1(i)}\rangle_{s_i}\big)  = (0,0,0).
\label{extconstsp}
\end{eqnarray}
All of the above-mentioned operators act in a Hilbert space $\mathcal{H}_{tot}$ $=$ $\otimes_{i=1}^3 \mathcal{H}^{(i)}_{tot} $ with an inner product  of vectors depending an all of the oscillators and ghosts, $(a^{(i)},b^{(i)};\eta^{(i)}_0,\eta^{(i)}_1, \mathcal{P}^{(i)}_1,\eta^{(i)}_{11}, \mathcal{P}^{(i)}_{11})\equiv (\mathcal{A}^{(i)}$; $\mathcal{C}^{(i)}$, $\mathcal{P}^{(i)})$  \cite{BRcub}:
\begin{eqnarray}
\hspace{-1.5ex}&\hspace{-1.5ex}&\hspace{-1.5ex} \langle\chi^{(i)} |\psi^{(j)}\rangle \hspace{-0.25ex}= \delta^{ij} \hspace{-0.25ex}\int\hspace{-0.3ex} d^d x \langle0|  \chi^{(i)*}\big(\mathcal{A}^{(i)};\mathcal{C}^{(i)},\mathcal{P}^{(i)}\big)\psi^{(j)}\big(\mathcal{A}^{(i)+};\mathcal{C}^{(i)+},\mathcal{P}^{(i)+}\big)|0\rangle\hspace{-0.2ex}.
\label{scalarprod}
\end{eqnarray}

The complete BRST operator $Q^{tot} = \sum_{j=1}^3Q^{(j)}$ supercommutes with any of $\sigma^{(i)}$; it is nilpotent in a subspace with zero $\hspace{ -0.1em}$ eigenvectors for the spin operators $\sigma^{(i)}$ (\ref{extconstsp}) and is Hermitian together with the operator $K = \otimes_{j=1}^3K^{(j)} $ with respect to the inner product (\ref{scalarprod}):
  \begin{align}\label{geneq}
   & (Q^{tot})^2 =  \sum_{i=1}^3\eta_{11}^{(i)+}\eta^{(i)}_{11} \sigma^{(i)} ,\qquad   Q^{tot+}K =  KQ^{tot} ; \\
     &    K= \otimes_{j=1}^3 \sum_{n_j=0}^{\infty}\frac{1}{n_j!}(b^{(j)+})^{n_j}|0\rangle\langle 0|(b^{(j)})^n
   \prod_{i_j=0}^{n_j-1}(i_j+h^{(j)}(s_j)).
   \label{geneq2}
  \end{align}

The vertex $ \big| V^{(3)}\rangle_{(s)_{3}}$ has a local representation:
\begin{equation}\label{xdep}
  \big |V^{(3)}\rangle_{(s)_3} = \prod_{i=2}^3 \delta^{(d)}\big(x_{1} -  x_{i}\big) V^{(3)}
  \prod_{j=1}^3 \eta^{(j)}_0 |0\rangle , \ \  |0\rangle\equiv \otimes_{e=1}^3 |0\rangle^{e}.
\end{equation}
The vertex is a BRST-closed solution of the equations \cite{BRcub}:
\begin{equation}
\label{g1Lmod}
    Q^{tot}
\big|{V}{}^{(3)}\rangle_{(s)_3}  =0, \ \ \sigma^{(i)}\big|{V}{}^{(3)}\rangle_{(s)_3}= 0,
\end{equation}
 (with the properties $(\epsilon, gh)\big|{V}{}^{(3)}\rangle = (1,3)$) as a consequence of the inner product completeness, as well as of the spin equations
 (\ref{extconstsp}). Arbitrariness in solutions of the system (\ref{g1Lmod}) is determined by adding BRST-exact terms of spin $(s)_3$,
\begin{equation}
\label{g1Lmodr}
    \big|\overline{V}{}^{(3)}\rangle_{(s)_3}  = \big|{V}{}^{(3)}\rangle_{(s)_3} + Q^{tot} \big|{X}{}^{(3)}\rangle_{(s)_3} , \ \ \sigma^{(i)}\big|{X}{}^{(3)}\rangle_{(s)_3}= 0,
\end{equation}
 ($(\epsilon, gh)\big|{X}{}^{(3)}\rangle = (0,2)$) which do not alter the equations of motion for the interacting model.

The gauge transformations form a closed algebra with a commutator of transformations being proportional to
the gauge transformation
\begin{eqnarray}
&&  \big[\delta^{\Lambda_1}_{[1]},\delta^{\Lambda_2}_{[1]}\big\}  |\chi^{(i)} \rangle \  =
\  - g \delta^{\Lambda_3}_{[1]}  |\chi^{(i)} \rangle  \
\label{closuregtr},
\end{eqnarray}
with a Grassmann-odd gauge parameter $\Lambda_3$, expressed functionally through $\Lambda_1$ and $\Lambda_2$: $\Lambda^{(i)}_3
 =$ $\Lambda^{(i)}_3(\Lambda_1, \Lambda_2)$. It should be noted that the validity of the Jacobi identity for the gauge transformation algebra imposes
 additional restrictions on the vertex $\big|{V}{}^{(3)}\rangle_{(s)_3}$.

The equation  (\ref{g1Lmod}) determines cubic interaction vertices for irreducible massless totally symmetric higher-spin fields.

We emphasize that a Lagrangian description without the interaction vertex $\big|  V^{(3)}\rangle_{(s)_{3}}$ is equivalent to 3 copies
of Fronsdal formulations \cite{Fronsdal} in terms of totally symmetric double traceless fields $\phi^{(i)}_{\mu(s_i)}$ and
traceless gauge parameters  $\xi^{(i)}_{\mu(s_i-1)}$, $i=1,2,3$.

\section{General solution for a cubic vertex: the form of the operator $\mathcal{Z}_j$}
\label{sec:BRSTinterZ}

A general solution of the equations (\ref{g1Lmod}) for a cubic vertex has been obtained \cite{BRcub} in the form of modified products
of special operators homogeneous in powers of oscillators (taking into account the conservation law for the momentum associated with the vertex)
\begin{eqnarray}\label{genvertex}
   \hspace{-1.0em} &\hspace{-1.0em}&\hspace{-1.0em}|{V}{}^{(3)}\rangle_{(s)_3} = |{V}{}^{M(3)}\rangle_{(s)_3}  +\hspace{-0.5em} \sum_{(j_1,j_2,j_3) >0}^{([s_{1}/2],[s_{2}/2],[s_{3}/2])}\hspace{-0.5em} U^{(s_1)}_{j_1}U^{(s_2)}_{j_2}U^{(s_3)}_{j_3}|{V}{}^{M(3)}\rangle_{(s)_3-2(j)_3}\hspace{-0.1ex},\\
  \label{Vmets}
 \hspace{-0.4ex} &\hspace{-0.4ex}&\hspace{-0.4ex} |{V}{}^{M(3)}\rangle_{{(s)_3-2(j)_3}}= \sum_{k}\mathcal{Z}_{1/2\{(s-2J) - k\}}\prod_{i=1}^3 \mathcal{L}^{(i)}_{s_{i}-2j_i-1/2(s-2J- k)} ,  \\
 \hspace{-0.4ex} &\hspace{-0.4ex}&\hspace{-0.4ex} (s,J) = \big(\sum_{i}s_i , \ \sum_ij_i\big). \nonumber
   \end{eqnarray}
The vertex $\big|{V}{}^{M(3)}\rangle_{(s)_3-2(j)_3}$ is defined in \cite{BRST-BV3} using the powers of operators linear $({L}^{(i)})^{k_i}$ and cubic  ${Z}^j$ in the powers of oscillators,
\begin{eqnarray}
  \label{LrZ}
  \hspace{-0.9em}&\hspace{-0.9em}&\hspace{-0.9em}  L^{(i)} \ = \   (p^{(i+1)}_{\mu}-p^{(i+2)}_{\mu})a^{(i)\mu+} - \imath \big(\mathcal{P}^{(i+1)}_0- \mathcal{P}^{(i+2)}_0
  \big)\eta_1^{(i)+} ,  \\
  \hspace{-0.9em}&\hspace{-0.9em}&\hspace{-0.9em} Z \ = \  L^{(12)+}_{11}L^{(3)} + L^{(23)+}_{11}L^{(1)} + L^{(31)+}_{11}L^{(2)}, \label{LrZ1}\\
  \hspace{-0.9em}&\hspace{-0.9em}&\hspace{-0.9em} L^{(i i+1)+}_{11} \ = \ \textstyle\frac{1}{2}a^{(i)\mu+}a^{(i+1)+}_{\mu} +
\frac{1}{2}\mathcal{P}^{(i)+}_1\eta_1^{(i+1)+} +
\frac{1}{2}\mathcal{P}^{(i+1)+}_1\eta_1^{(i)+}, \label{Lrr+1}
\end{eqnarray}
with a subsequent replacement by BRST $Q^{tot}$-closed forms
$\mathcal{L}^{(i)}_{k_i}$,  $k_i = 1, ..., s_i$ and operators $ \mathcal{Z}_j$.
Here,  $p^{(i)}_{\mu} = -i\partial^{(i)}_{\mu}$, and the quantities $\mathcal{L}^{(i)}_{k_i}$ are given by the rule
\begin{eqnarray}\label{LrZLL}
\hspace{-0.9em}&\hspace{-0.9em}&\hspace{-0.9em}  \mathcal{L}^{(i)}_{k_i} \ = \  ({L}^{(i)})^{k_i-2} \Big(({L}^{(i)})^{2} -  \frac{\imath k_i!}{2(k_i-2)!} \eta_{11}^{(i)+} \big[ 2\mathcal{P}^{(i+1)}_0+ 2 \mathcal{P}^{(i+2)}_0-\mathcal{P}^{(i)}_0
\big] \Big).
\end{eqnarray}
The set of  $Q^{tot}$- closed operators also includes new two-, four-, ..., $[s_i
/2]$ forms in powers of oscillators, corresponding to trace operators at $i=1,2,3$
\begin{equation}\label{trform}
U^{(s_i)}_{j_i}\big(\eta_{11}^{(i)+},\mathcal{P}_{11}^{(i)+} \big) \
: = \
(\widehat{L}{}^{(i)+}_{11})^{(j_i-2)}\big\{(\widehat{L}{}^{+(i)}_{11})^2
- j_i(j_i-1)\eta_{11}^{(i)+}\mathcal{P}_{11}^{(i)+}\big\}.
\end{equation}
Different representatives of vertices are labelled by a natural-valued parameter $k$, restricted by the inequalities
 \begin{eqnarray}
  s-2J-2s_{\min}\leq k\leq s-2J, \ \ \ k=s-2J - 2p, \ p\in \mathbb{N}_0.
\end{eqnarray}
 so that the order of derivatives diminishes in the representatives by the value of 2 under the change $k\to k+1$.
 Notice that the vertex $|{V}{}^{(3)}\rangle$ (\ref{genvertex}) may contain terms without derivatives for even-valued
 helicities $s_i$, as well as some terms with one, two, and three derivatives in case the respective one, two, and all
 the $s_i$ helicities are odd-valued \cite{BRcub}.

The quantity $\mathcal{Z}_j$ в (\ref{Vmets}) is defined in \cite{BRcub} for $j=1$ by the relation
\begin{eqnarray}
  \hspace{-0.6em}&\hspace{-0.6em}&\hspace{-0.6em} \mathcal{Z}_1\prod_{i=1}^3\mathcal{L}^{(i)}_{k_i} = {Z}\prod_{i=1}^3\mathcal{L}^{(i)}_{k_i} - \sum_{l=1}^3k_l\frac{b^{(l)+}}{h^{(l)}}\Big[\Big[\widehat{L}{}^{(l)}_{11},{Z}\Big\},{L}^{(l)} \Big\}\prod_{i=1}^3\mathcal{L}^{(i)}_{k_i-\delta_{il}} \label{calZ}\\
 \hspace{-0.6em}&\hspace{-0.6em}&\hspace{-0.6em}  +\sum_{l\ne e}^3 k_lk_e\frac{b^{(l)+}b^{(e)+}}{h^{(l)}h^{(e)}}\Big[\widehat{L}{}^{(e)}_{11},\Big[\Big[\widehat{L}{}^{(l)}_{11},{Z}\Big\},{L}^{(l)} \Big\}{L}^{(e)} \Big\}\prod_{i=1}^3\mathcal{L}^{(i)}_{k_i-\delta_{il}-\delta_{ei}} \nonumber\\
 \hspace{-0.7em}&\hspace{-0.7em}&\hspace{-0.7em}  - \sum_{l\ne e\ne o}^3k_lk_ek_o\frac{b^{(l)+}b^{(e)+}b^{(o)+}}{h^{(l)}h^{(e)}h^{(o)}}\Big[\widehat{L}{}^{(o)}_{11},\Big[\widehat{L}{}^{(e)}_{11},\Big[\Big[\widehat{L}{}^{(l)}_{11},{Z}\Big\},{L}^{(l)} \Big\} {L}^{(e)} \Big\}{L}^{(0)} \Big\}\prod_{i=1}^3\mathcal{L}^{(i)}_{k_i-1}. \nonumber
\end{eqnarray}
The $Q^{tot}$-closeness of $\mathcal{Z}_1\prod_{i=1}^3\mathcal{L}^{(i)}_{k_i}$ follows from the $Q^{tot}$-closeness of $\mathcal{L}^{(i)}_{k_i}$
and also from the fact that the trace-dependent part of the BRST operator
$\eta^{(l)+}_{11}\widehat{L}{}^{(l)}_{11}$, being the only source of the failure of the operator $Z$ (\ref{LrZ1}) to be BRST-closed,
transforms the initial operator into a product of the quantity $\Big[\Big[\widehat{L}{}^{(l)}_{11},{Z}\Big\},{L}^{(l)} \Big\}$-independent of
 the oscillators carrying the index $l$ and the $Q^{tot}$-closed form $\prod_{i=1}^3\mathcal{L}^{(i)}_{k_i-\delta_{il}}$.
Under an additive subtraction of the indicated product multiplied by $k_l\frac{b^{(l)+}}{h^{(l)}}$, respectively, for each $l=1,2,3$ from the initial value,
one obtains
\begin{eqnarray}\label{prZ1}
&& Q^{tot}\Big( {Z}\prod_{i=1}^3\mathcal{L}^{(i)}_{k_i} - \sum_{l=1}^3k_l\frac{b^{(l)+}}{h^{(l)}}\Big[\Big[\widehat{L}{}^{(l)}_{11},{Z}\Big\},{L}^{(l)} \Big\}\prod_{i=1}^3\mathcal{L}^{(i)}_{k_i-\delta_{il}} \Big)  \\
&&  = \sum_{l=1}^3\eta^{(l)+}_{11}k_l\bigg\{\Big[\Big[\widehat{L}{}^{(l)}_{11},{Z}\Big\},{L}^{(l)} \Big\}\nonumber\\
&& \phantom{=} - \Big(h^{(l)}b^{(l)}+\sum_{e\ne l}^3\eta^{(e)+}_{11}\widehat{L}{}^{(e)}_{11}\Big)\frac{b^{(l)+}}{h^{(l)}}\Big[\Big[\widehat{L}{}^{(l)}_{11},{Z}\Big\},{L}^{(l)}  \bigg\}\prod_{i=1}^3\mathcal{L}^{(i)}_{k_i-\delta_{il}}\nonumber
\\
&& = - \sum_{l,e\ne l}^3 \eta^{(e)+}_{11}k_lk_e\frac{b^{(l)+}}{h^{(l)}}\Big[\widehat{L}{}^{(e)}_{11},\Big[\Big[\widehat{L}{}^{(l)}_{11},{Z}\Big\},{L}^{(l)} \Big\}{L}^{(e)} \Big\}\prod_{i=1}^3\mathcal{L}^{(i)}_{k_i-\delta_{il}-\delta_{ei}}. \nonumber
\end{eqnarray}
In (\ref{prZ1}), we take into account (\ref{L11L11+}) that it is only the part $h^{(l)}b^{(l)}$ of the operator $L^{(l)}_{11}$ that acts non-trivially
on the second term. As we notice, once again, that the structure of the final expression in (\ref{prZ1}) consists of a $Q^{tot}$-closed part and
a triple supercommutator, independent of the oscilltors carrying the indices $l,e $, with $l\ne e$, except for the ``processed'' oscillator $b^{(l)+}$,
we introduce additively the indicated product for each $e\ne l$, $ e=1,2,3$, multiplied, respectively, by $k_e\frac{b^{(e)+}}{h^{(e)}}$, thereby increasing
the first two terms (\ref{calZ}).

As a result, under the action of $Q^{tot}$ on a twice-modified quantity, we obtain
\begin{eqnarray}\label{prZ11}
&& Q^{tot}\Big( {Z}\prod_{i=1}^3\mathcal{L}^{(i)}_{k_i} - \sum_{l=1}^3k_l\frac{b^{(l)+}}{h^{(l)}}\Big[\Big[\widehat{L}{}^{(l)}_{11},{Z}\Big\},{L}^{(l)} \Big\}\prod_{i=1}^3\mathcal{L}^{(i)}_{k_i-\delta_{il}}\\
&& +\sum_{l\ne e}^3 k_lk_e\frac{b^{(l)+}b^{(e)+}}{h^{(l)}h^{(e)}}\Big[\widehat{L}{}^{(e)}_{11},\Big[\Big[\widehat{L}{}^{(l)}_{11},{Z}\Big\},{L}^{(l)} \Big\}{L}^{(e)} \Big\}\prod_{i=1}^3\mathcal{L}^{(i)}_{k_i-\delta_{il}-\delta_{ei}} \Big) = \nonumber \\
\hspace{-0.75em}&\hspace{-0.75em}&\hspace{-0.75em}   \sum_{l\ne e\ne o}^3\eta^{(o)+}_{11}k_lk_ek_o\frac{b^{(l)+}b^{(e)+}}{h^{(l)}h^{(e)}}\Big[\widehat{L}{}^{(o)}_{11},\Big[\widehat{L}{}^{(e)}_{11},\Big[\Big[\widehat{L}{}^{(l)}_{11},{Z}\Big\},{L}^{(l)} \Big\} {L}^{(e)} \Big\}{L}^{(0)} \Big\}\prod_{i=1}^3\mathcal{L}^{(i)}_{k_i-1}.\nonumber
\end{eqnarray}
The structure of the non-vanishing expression in (\ref{prZ11}) consists, once again, of a $Q^{tot}$-closed part and the fourth supercommutator, independent of all of the oscillators, except the ``processed'' ones, $b^{(l)+}$, $b^{(e)+}$. Subtracting the latter term, constructed as multiplying respectively by  $k_o\frac{b^{(o)+}}{h^{(o)}}$, from the first three terms in (\ref{calZ}) proves the BRST-closeness of the quantity (\ref{calZ}).

For $j=2$, we repeat the suggested algorithm, starting from $Z \mathcal{Z}\times$ $\times\prod_{p=1}^3\mathcal{L}^{(p)}_{k_p}$, and finally obtain
\begin{eqnarray}
  && \mathcal{Z}_2\prod_{i=1}^3\mathcal{L}^{(i)}_{k_i} = Z\mathcal{Z}\prod_{i=1}^3\mathcal{L}^{(i)}_{k_i}- \sum_{i_1=1}^3\frac{b^{(i_1)+}}{h^{(i_1)}}\Big[\Big[\widehat{L}{}^{(i_1)}_{11},{Z}\Big\}, \mathcal{Z}\prod_{i=1}^3\mathcal{L}^{(i)}_{k_i} \Big\} \label{calZZ}\\
&& +\sum_{i_1\ne e_1}^3 \frac{b^{(i_1)+}b^{(e_1)+}}{h^{(i_1)}h^{(e_1)}}\Big[\widehat{L}{}^{(e_1)}_{11},\Big[\Big[\widehat{L}{}^{(i_1)}_{11},{Z}\Big\},\mathcal{Z}\prod_{i=1}^3\mathcal{L}^{(i)}_{k_i}\Big\}\Big\} \nonumber\\
&& - \sum_{i_1\ne e_1\ne o_1}^3\frac{b^{(i_1)+}b^{(e_1)+}b^{(o_1)+}}{h^{(i_1)}h^{(e_1)}h^{(o_1)}}\Big[\widehat{L}{}^{(o_1)}_{11},\Big[\widehat{L}{}^{(e_1)}_{11},
\Big[\Big[\widehat{L}{}^{(i_1)}_{11},{Z}\Big\},\mathcal{Z}\prod_{i=1}^3\mathcal{L}^{(i)}_{k_i}
\Big\}\Big\}\Big\}. \nonumber
\end{eqnarray}
For $j+1 \geq 1$, in turn, we obtain by induction
\begin{eqnarray}
  && \mathcal{Z}_{j+1}\prod_{i=1}^3\mathcal{L}^{(i)}_{k_i} = Z\mathcal{Z}_{j}\prod_{i=1}^3\mathcal{L}^{(i)}_{k_i} - \sum_{i_{j}=1}^3\frac{b^{(i_{j})+}}{h^{(i_{j})}}\Big[\Big[\widehat{L}{}^{(i_{j})}_{11},{Z}\Big\}, \mathcal{Z}_{j}\prod_{i=1}^3\mathcal{L}^{(i)}_{k_i} \Big\} \label{calZj}\\
&& +\sum_{i_{j}\ne e_{j}}^3 \frac{b^{(i_{j})+}b^{(e_{j})+}}{h^{(i_{j})}h^{(e_{j})}}\Big[\widehat{L}{}^{(e_{j})}_{11},
\Big[\Big[\widehat{L}{}^{(i_{j})}_{11},{Z}\Big\},\mathcal{Z}_{j}\prod_{i=1}^3\mathcal{L}^{(i)}_{k_i}\Big\}\Big\} \nonumber\\
&& - \hspace{-0.5em} \sum_{i_{j}\ne e_{j}\ne o_{j}}^3\hspace{-0.5em}\frac{b^{(i_{j})+}b^{(e_{j})+}b^{(o_{j})+}}{h^{(i_{j})}h^{(e_{j})}h^{(o_{j})}}\Big[\widehat{L}{}^{(o_{j})}_{11},
\Big[\widehat{L}{}^{(e_{j})}_{11},\Big[\Big[\widehat{L}{}^{(i_{j})}_{11},{Z}\Big\},\mathcal{Z}_{j}\prod_{i=1}^3\mathcal{L}^{(i)}_{k_i}
\Big\}\Big\}\Big\}. \nonumber
\end{eqnarray}
The relations (\ref{calZ}), (\ref{calZZ}),  (\ref{calZj}) determine the quantities $\mathcal{Z}_j$ in the cubic vertex (\ref{genvertex}), which
presents the main result of this paper.

\section{Conclusion}
\label{sec:conclusion}

In the present article, we have obtained an exact representation for the quantities $\mathcal{Z}_j$, for $j\geq 1$, that constitute the non-multiplicative
part of a general cubic vertex constructed in \cite{BRcub} for massless completely symmetric fields of arbitrary integer helicities
$s_1$, $s_2$, $s_3$ in a $d$-dimensional Minkowski spacetime.

The construction is implemented in the framework of an unconstrained BRST approach to higher-spin field theory,
in which every condition that determines an irreducible massless representation of higher spin
is taken into account on an equal footing in the complete BRST operator, as compared to all the previous studies.
As a consequence, the cubic Lagrangian vertex operator (\ref{genvertex}) preserves both the locality and  the irreducibility property
of a representation for interacting fields of helicities $s_1$, $s_2$, $s_3$.

The inclusion of trace restrictions into the BRST operator has led to a larger content of configuration
spaces in Lagrangian formulations for interacting fields of integer helicities in question
(as compared to the constrained BRST approach \cite{BRST-BV3}), which has permitted the appearance
of new trace operator components $U^{(s_i)}_{j_i}$ (\ref{trform}) in the cubic vertex.
In this regard, the correspondence between the obtained vertex $|{V}{}^{(3)}\rangle$ and the vertex $|{V}{}^{M(3)}\rangle$ of \cite{BRST-BV3} is not unique due to the fact that the tracelessness conditions for the latter vertex are not satisfied: ${L}{}^{(i)}_{11}\times$ $|{V}{}^{M(3)}\rangle \ne 0$. Both vertices will correspond to each other, firstly, after extracting the irreducible components $|{V}{}^{M(3)}_{irrep}\rangle$ from $|{V}{}^{M(3) }\rangle$, satisfying ${L}{}^{(i)}_{11}|{V}{}^{M(3)}_{irrep}\rangle = 0$. Secondly, after eliminating the auxiliary fields and gauge parameters by partially fixing the gauge and using the equations of motion, the vertex $|{V}{}^{(3)}\rangle $ will transform to $|{\breve{V}}{}^{(3)}\rangle$ in a triplet formulation of \cite{BRST-BV3}, so that, up to total derivatives, the vertices $|{V}{}^{M(3)}_{irrep}\rangle$ and $|{\breve{V}}{}^{(3)}\rangle$ must coincide.

The construction of an irreducible cubic vertex $|{V}{}^{M(3)}_{irrep}\rangle$ poses an interesting problem.
The suggested approach can be further developed: for irreducible massless half-integer higher-spin fields on a flat
background; for massive integer and half-integer higher-spin fields; for higher-spin fields of a mixed index symmetry;
for supersymmetric fields of higher spins, where the vertices must include any degree of traces. One should also mention
the problem of constructing the quartic and higher vertices in the BRST approach, as well as the quantization of
a model of interacting higher-spin fields, by following the algorithm for constructing a quantum BRST--BV action
\cite{BurdikResh}. All the mentioned problems are awaiting their solution in our forthcoming works.

\paragraph{Acknowledgements} The work has been carried out under  the Ministry of Education of Russian
Federation, project No. FEWF-2020-003. The author is grateful to I.L.~Buchbinder for helpful discussions of the presented results.


\end{document}